\documentclass[aps,preprint,onecolumn,superscriptaddress,nofootinbib]{revtex4-1}

\usepackage{graphicx}
\usepackage{dcolumn}
\usepackage{bm}

\begin{document}


\title{Metal-free Stoner and Mott-Hubbard magnetism in 2D polymers with honeycomb lattice}%

\author{Hongde Yu}
\affiliation{Faculty of Chemistry and Food Chemistry, TU Dresden, Bergstrasse 66c, 01069 Dresden, Germany}

\author{Thomas Heine}
    \email{thomas.heine@tu-dresden.de}
    \affiliation{Faculty of Chemistry and Food Chemistry, TU Dresden, Bergstrasse 66c, 01069 Dresden, Germany}
    \affiliation{Helmholtz-Zentrum Dresden-Rossendorf, Institute of Resource Ecology, Permoserstrasse 15, 04318 Leipzig, Germany}
    \affiliation{Department of Chemistry, Yonsei University, Seodaemun-gu, Seoul 120-749, Republic of Korea}

\date{\today}

\begin{abstract}
We computationally demonstrate Stoner-ferromagnetic half-metals and antiferromagnetic Mott-Hubbard insulators in metal-free 2D polymers. Coupling radicaloid (hetero)triangulene monomers via strong covalent bonds preserving the in-plane conjugation of the electronic $\pi$ system yields 2D crystals with long-range magnetic order and magnetic couplings above the Landauer limit. Dual-site honeycomb lattices produce both flat bands and Dirac cones. Depending on the monomers, electron correlations lead to either a bandgap at the Dirac points for antiferromagnetic Mott insulators, or Stoner ferromagnetism with both spin-polarized Dirac cones and flat bands at the Fermi level. These results pioneer a new type of Stoner and Mott-Hubbard magnetism emerging in the electronic $\pi$ system of crystalline conjugated 2D polymers.
\end{abstract}

\maketitle


Long-range magnetic ordering in metal-free materials is one of the unicorns of materials science. While early investigations into magnetic carbon derived from pressurized fullerenes garnered considerable interest, they also raised skepticism; subsequent studies attributed observed phenomena to impurities or defects \cite{Makarova.2001}. The foundation of metal-free magnetism lies in both stable paramagnetic centers and the ability to control magnetic coupling. Half-filled flat bands in two-dimensional (2D) materials, such as magic-angle twisted bilayer graphene \cite{Bultinck.2020,Sharpe.2021,Tschirhart.2021}, zigzag carbon nanoribbons \cite{Son.2006,Fujita.1996} and 2D polymers with Lieb or kagome lattice \cite{Springer.2020,Thomas.2019,Jiang.2019,Jin.2017,Galeotti2020}, offer the exploitation of structural topology to generate paramagnetic centers, partially with long-range ordering. These spin-polarized flat bands harbor a variety of exotic quantum phenomena, ranging from fractional quantum Hall effect \cite{Tang.2011,Neupert.2011}, unconventional superconductivity \cite{Cao.2018,Balents.2020,Peri.2021}, non-Fermi liquid \cite{Sayyad.2020} to high-order topological insulators \cite{Ni.2022}, spin frustration \cite{Alcon.2023}, and Mott insulators \cite{Thomas.2019,Wu.2018}. However, controlling the magnetic coupling in such systems remains particularly challenging, especially above cryogenic conditions and when aiming for metal-free ferromagnetism (FM). In this letter, we propose a novel design concept for 2D polymers, encompassing diverse magnetic orderings and exhibiting robust coupling constants. The concept is based on stable paramagnetic building blocks that are coupled via covalent bonds to form 2D polymers with strong in-plane conjugation. Leveraging strategic lattice topology design, we observed both spin-polarized linear band crossings and flat bands.

\begin{figure}
\includegraphics[width=76mm]{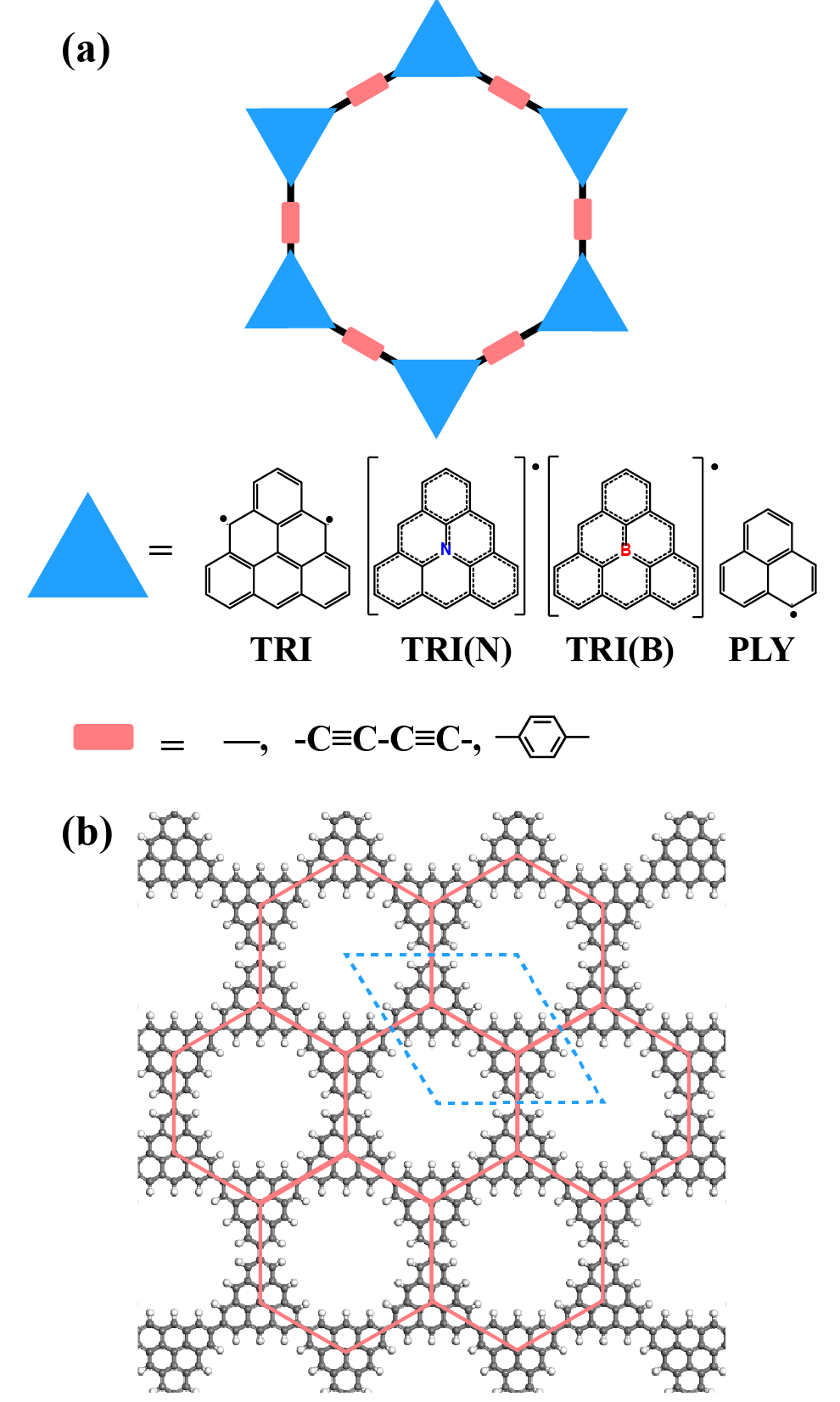}
\caption{\label{fig:fig1} Chemical structures of (hetero)triangulene monomers and linkages (a) and 2D polymer (b). The red line in b shows the honeycomb lattice and the blue line illustrates the unit cell.}
\end{figure}

\begin{figure}
\includegraphics[width=76mm]{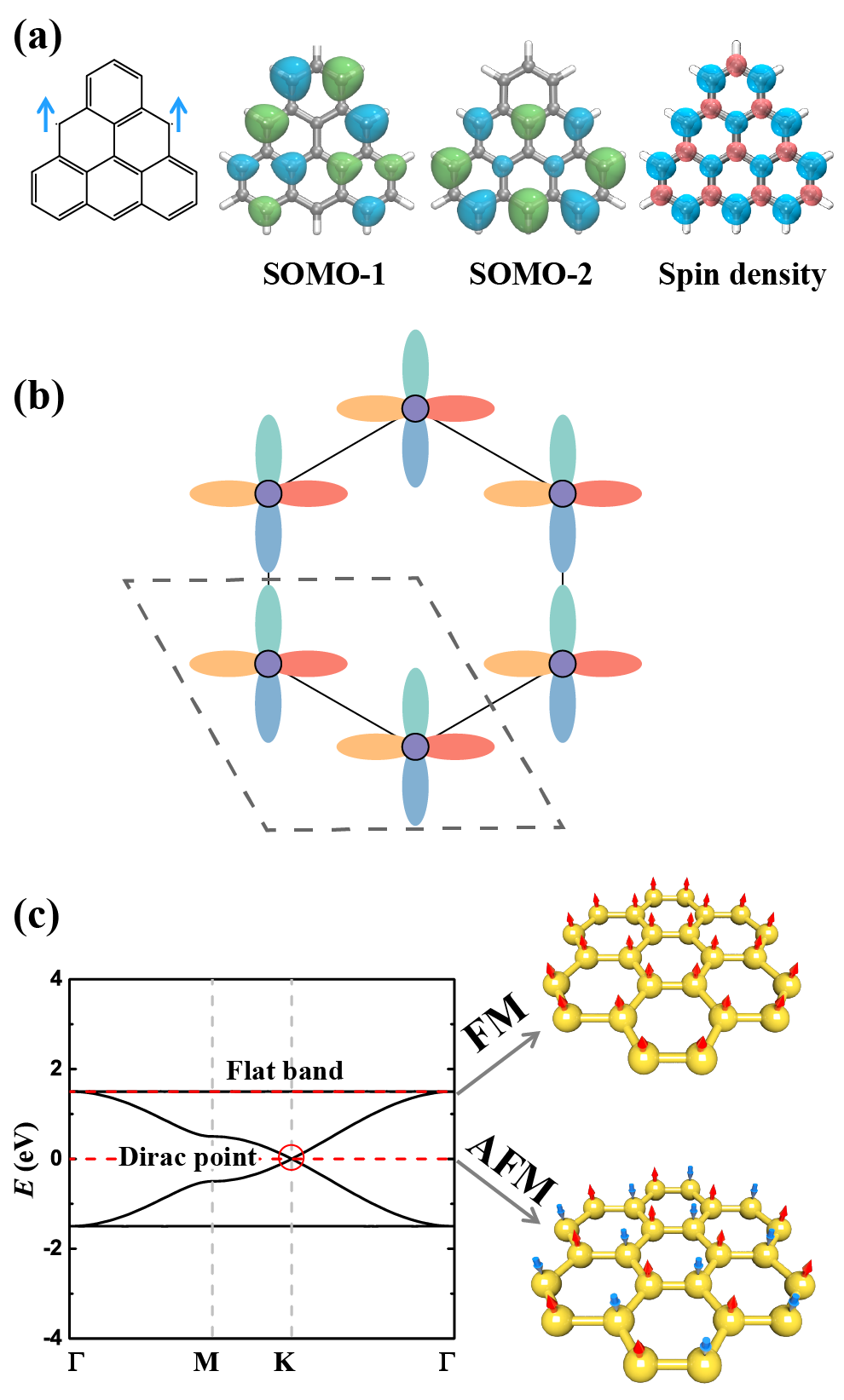}
\caption{\label{fig:fig2} (a) Chemical structure, the singly occupied molecular orbitals (SOMOs) and the spin density distribution of triangulene monomer. Illustrations of honeycomb 2D polymer with a ($\pi_x, \pi_y$) basis (b) and corresponding band structure (c) with Dirac point flanked by two flat bands. Upon tuning the electron filling, the systems potentially change from antiferromagnetic (AFM) to ferromagnetic (FM).}
\end{figure}

Triangulene (TRI), first proposed by Erich Clar in 1953 \cite{Clar.1953,Clar.1964}, is recognized as the smallest nanographene featuring a triplet ground state. This biradical with a unique non-Kekulé structure has been synthesized both in solution and on the surface (Figs.~\ref{fig:fig1}a) \cite{Mishra.2020,Pavlicek.2017,Arikawa.2021}. In addition to pristine carbon-centered triangulene, hetero-triangulenes, such as nitrogen-centered triangulene (TRI(N)) \cite{Wang.2022,Wei.2022} and boron-centered triangulene (TRI(B)) \cite{Hirai.2019}, and smaller variant phenalenyl (PLY) (i.e., [2] triangulene) \cite{Goto.1999,krane2023exchange,turco2023observation}, have also been realized, owing to rapid advancements in precision synthesis of planar organic radicals (Fig.~\ref{fig:fig1}). Notably, the TRI monomer possesses two energy-degenerate orthogonal radical orbitals (i.e. singly occupied molecular orbitals (SOMOs)) (Fig.~\ref{fig:fig2}a) \cite{Pavlicek.2017}, differing from many mono-radicals with a single SOMO, such as PLY, perchlorotriphenylmethyl (PTM), triarylmethyl (TAM), and trioxotriangulene (TOT) \cite{Thomas.2019,Wu.2018}. These triangular-shaped monomers can be integrated into 2D polymers via well-established Yamamoto or Ullmann coupling reactions \cite{Mishra.2020,Mishra.2021,Niu.2022,Liu.2015}. Here, we create 15 novel 2D polymers, using these radicals as building blocks, linked directly or bridged by common linear linkages, including -C$\equiv$C-C$\equiv$C- (CCCC) and phenyl (Ph) (Figs.~\ref{fig:fig1} and S1-S5).

As shown in Fig.~\ref{fig:fig1}b, 2D polymers made of triangulene present a honeycomb lattice similar to graphene \cite{Novoselov.2004} (Fig.~\ref{fig:fig1}b and S6). However, unlike graphene and many mono-radical-based 2D polymers with only one half-filled orbital per lattice site, such as PLY, TOT, TAM, and PTM \cite{Thomas.2019,Novoselov.2004,Neto.2009}, the triangulene-based 2D polymers host two orthogonal orbitals at each site (denoted as $\pi_x$ and  $\pi_y$), originating from the degenerate SOMO orbitals of its triplet monomer (Fig.~\ref{fig:fig2}a). We first explore the electronic structure of this four-orbital-based 2D lattice by employing a tight-binding (TB) model with a ($\pi_x$, $\pi_y$) basis and the Slater-Koster scheme \cite{Slater.1954}. The corresponding effective Hamiltonian representing single-electron properties is expressed as
$H_{eff}=\left(\begin{array}{cc}
\begin{array}{cc}
\varepsilon_0 & 0 \\
0 & \varepsilon_0 \\
\end{array} & \begin{array}{cc}
V_{xx} & V_{xy} \\
V_{xy} & V_{yy} \\
\end{array} \\
\begin{array}{cc}
V_{xx}^\ast & V_{xy}^\ast \\
V_{xy}^\ast & V_{yy}^\ast \\
\end{array} & \begin{array}{cc}
\varepsilon_0 & 0 \\
0 & \varepsilon_0 \\
\end{array}
\end{array}\right)$,
where $V_{xx}=(\frac{3}{4}pp\sigma+\frac{1}{4}pp\pi)(e^{i\mathit{k}\mathit{a}_1}+e^{i\mathit{k}\mathit{a}_2})+pp\pi\ e^{i\mathit{k}(\mathit{a}_1+\mathit{a}_2)}$,
$\ V_{yy}=(\frac{1}{4}pp\sigma+\frac{3}{4}pp\pi)(e^{i\mathit{k}\mathit{a}_1}+e^{i\mathit{k}\mathit{a}_2})+pp{\sigma\ e}^{i\mathit{k}(\mathit{a}_1+\mathit{a}_2)}$, 
$V_{xy}=\frac{\sqrt3}{4}(pp\sigma-pp\pi)(e^{i\mathit{k}\mathit{a}_1}-e^{i\mathit{k}\mathit{a}_2})$, 
$pp\sigma$ and $pp\pi$ are the Slater-Koster parameters \cite{Ni.2022,Jiang.2021}.

As illustrated by Fig. ~\ref{fig:fig2}b and ~\ref{fig:fig2}c, the ($\pi_x$, $\pi_y$)-TB model, representing the 2D polymers of triangulenes with doubly degenerate frontier orbitals, yields a distinctive four-band structure: a Dirac point flanked by two flat bands. In contrast, graphene only shows a Dirac point near the Fermi level \cite{Neto.2009}. This four-orbital honeycomb lattice is different from many typical hexagonal lattices, such as one-orbital honeycomb, kagome and honeycomb-kagome lattices \cite{Springer.2020,Thomas.2019b}  (Fig. S7-S9). Such an unconventional electronic structure stems from the number of on-site orbitals and has seldomly been observed in mono-radical-based 2D polymers, such as those made of TOT, TAM and PTM \cite{Thomas.2019} (Fig. S10-S12). Beyond triangulene-based 2D polymers, this ($\pi_x$, $\pi_y$)-TB model can also represent many 2D polymers structured with a honeycomb lattice and comprised of monomers with doubly degenerate orbitals. For instance, it can delineate the band structure of diamagnetic 2D covalent organic framework (COF) with benzene nodes that also possesses energy-degenerate highest occupied molecular orbitals (HOMOs) \cite{Ni.2022} (Fig. S13).

\begin{figure*}
\includegraphics[width=\textwidth]{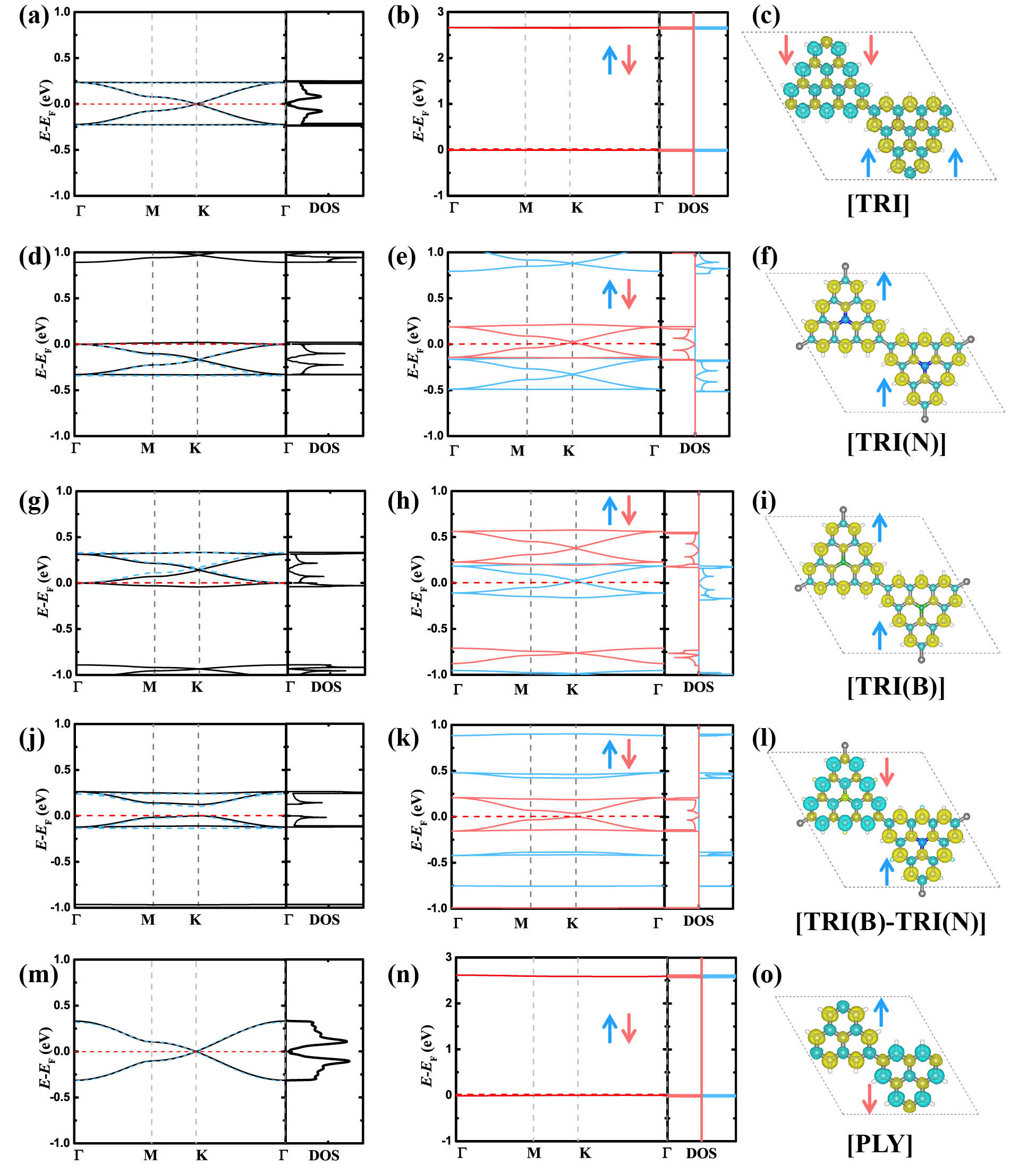}
\caption{\label{fig:fig3}Band structures and density of states of diamagnetic (a, d, g, j, m) and spin-polarized ground states (b, e, h, k, n) and spin density distributions (c, f, i, l, o) of triangulene-based 2D polymers, including [TRI] (a, b, c), [TRI(N)] (d, e, f), [TRI(B)] (g, h, i), [TRI(B)-TRI(N)] (j, k, l), and [PLY] (m, n, o). Blue dash lines in diamagnetic band structures are calculated by the tight-binding model. For [TRI], [PLY], [TRI(B)-TRI(N)], the ground states are AFM states, while they are FM states for [TRI(N)] and [TRI(B)].}
\end{figure*}

Under half-filling conditions, the Fermi level of carbon-centered triangulene 2D polymer, [TRI], intersects the Dirac point resembling graphene (Fig. ~\ref{fig:fig2}c and ~\ref{fig:fig3}a). Conversely, by adjusting the electron number of the system, it is feasible to shift the Fermi level to align with the flat band, for example, via introducing nitrogen or boron in [TRI(N)] and [TRI(B)]. Within the flat bands, the many-body effects typically dominate over the kinetic energy, resulting in a high state density of strongly-correlated electrons due to the interference effect, despite the presence of substantial orbital overlap (Fig.~\ref{fig:fig2}c and ~\ref{fig:fig3}d). This differs from the atomic flat bands observed at the core level or in the presence of dangling bonds, which originate from the lack of overlap between adjacent atomic wavefunction overlap \cite{Regnault.2022}. These flat bands can provoke many intriguing physical phenomena, such as fractional quantum Hall effects and unconventional superconductivity, as observed in twisted bilayer graphene and 2D kagome lattices \cite{Cao.2018,Sheng.2011,Torma.2022,Yankowitz.2019}. In addition, the high density of states near the Fermi level corresponding to the flat bands could induce spontaneous spin-polarization according to the Stoner model \cite{Coey.2010}, where $U/t > 1 $  is a critical value for phase transition. However, in most metal-free systems, introducing spin-polarization is of fundamental challenge due to the absence of degenerate orbitals coupled with the presence of strong electronic coupling ($t$) in $\pi$-orbitals. This contrasts sharply with inorganic compounds featuring transition metals and energy-degenerate d-orbitals

As carbon-centered triangulene 2D polymers, including [TRI] and [PLY], possess balanced sublattices, Ovchinnikov’s rule is applicable and predicts that they have a singlet ground state, i.e. $S = 0$ (Fig. S14-S16). However, it is unclear whether it is a closed-shell (diamagnetic) or open-shell (antiferromagnetic, AFM) singlet state. In order to understand the ground state and the spin-polarization in these polymers, we employed first-principles calculations as implemented in Vienna Ab initio Simulation Package \cite{Kresse.1996} (VASP 5.4.4) and CRYSTAL17 \cite{Dovesi.2018} (see computational details for more information). As illustrated in Fig.~\ref{fig:fig3}a, for the diamagnetic state, [TRI] shows a band structure featuring a Dirac point and twin flat bands, which can be well-represented by the four-orbital TB model with $t = 0.16$ eV. As in graphene, its Fermi level intersects the Dirac point (Fig.~\ref{fig:fig3}a). However, unlike graphene's diamagnetic and semi-metallic ground state, the spin-polarized singlet state in [TRI] is 2.17 eV more stable than the diamagnetic one. This is in accord with the non-Kekulé structure and corresponding unpaired electrons as featured by its radical building blocks. This is further manifested by a substantial $U/t$ ratio of 16.6 as per the Stoner model (Table ~\ref{tab:table1}). Hereafter, to evaluate the stability of spin-polarized states, the spin-polarization energy ($\Delta E_{\text{spin}}$) is defined by the energy difference between spin-polarized ground state ($E_{\text{GS}}$) and diamagnetic state ($E_{\text{CS}}$), $\Delta E_{\text{spin}} = E_{\text{GS}} - E_{\text{CS}}$.

\begin{figure}
\includegraphics[width=76mm]{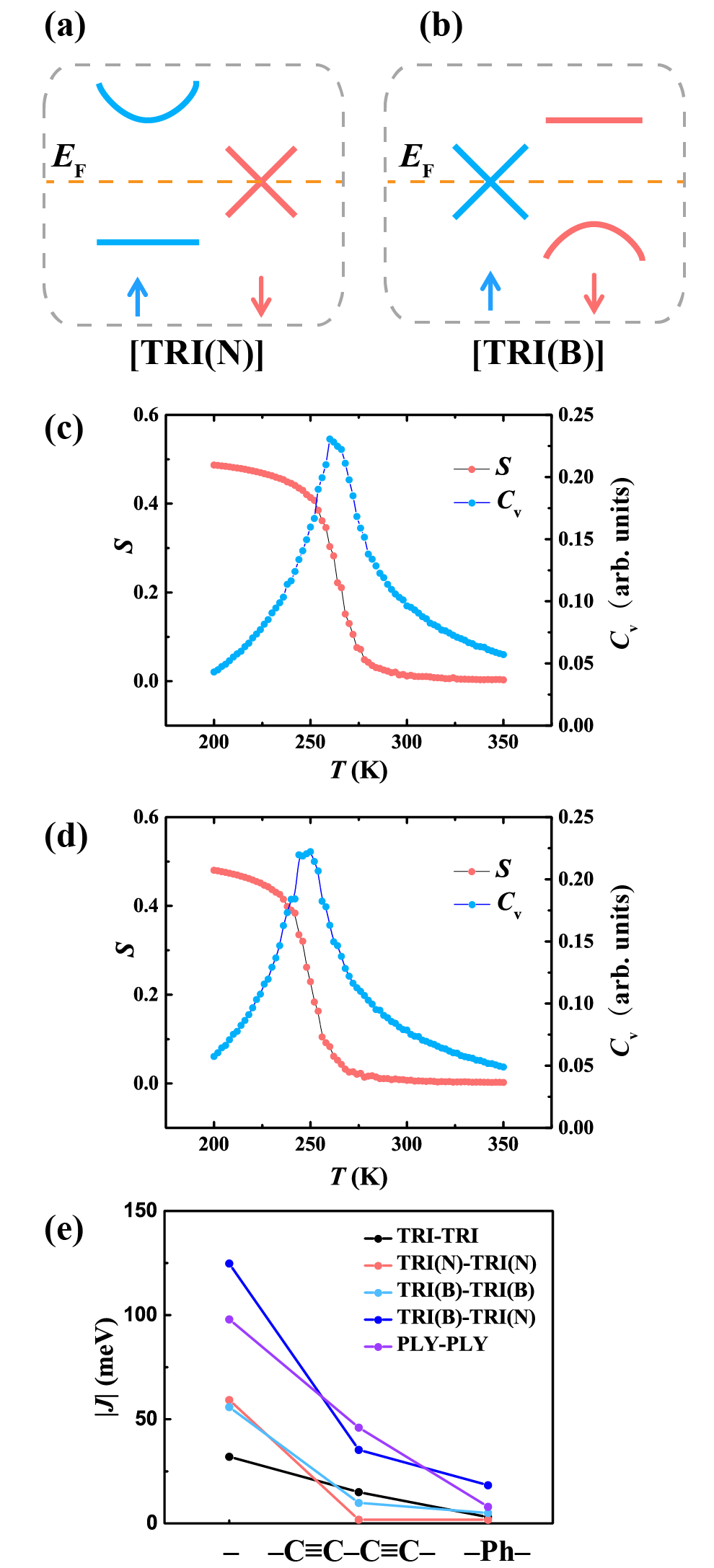}
\caption{\label{fig:fig4}  Illustration of half-metallic band structures of [TRI(N)] (a) and [TRI(B)] (b). Normalized spin ($S$) per site and electronic heat capacity ($C_{\text{v}}$) with the temperature from the Monte Carlo (MC) simulations of [TRI(N)] (c) and [TRI(B)] (d). (e) Correlations between linkages and strength of magnetic coupling ($|J|$).}
\end{figure}

We further find for the [TRI] that the AFM state is more stable than the FM state, with a magnetic coupling of -32 meV between adjacent sites, according to the Heisenberg–Dirac–van Vleck (HDVV) Hamiltonian, $\hat{H}=-\sum_{<i,\ j>}{{J\hat{S}}_i{\hat{S}}_j}$ (Fig. ~\ref{fig:fig3}b, ~\ref{fig:fig3}c and Table ~\ref{tab:table1}). From a spin-exchange perspective, the strong electronic coupling of 0.16 eV between neighbouring sites results in a predominant super-exchange (kinetic exchange) of -38 meV, favoring AFM interaction and anti-parallel spin alignment (Table ~\ref{tab:table1}). As a consequence of the pronounced on-site Coulomb repulsion, [TRI] becomes a Mott insulator with a bandgap of 2.66 eV, thus diverging from the “gapless” semi-metallic nature of graphene (Fig.~\ref{fig:fig3}b). Indicated by the degenerate AFM bands of [TRI], the spin-up and spin-down electrons are energetically equivalent and delocalized in momentum space, whereas, in real space, they are highly localized within adjacent monomers owning to the on-site Coulomb repulsion (Fig. ~\ref{fig:fig3}b and ~\ref{fig:fig3}c). Similar phenomena are also observed in metal-oxide complexes and 2D polymers made of mono-radicals, such as PTM-COF, TOT-COF, and TAM-COF \cite{Coey.2010,Wu.2018,Thomas.2019}. Analogously, other half-filled 2D polymers, including [PLY] and [TRI(B)-TRI(N)], also obey Ovchinnikov’s rule and exhibit AFM ground states with $J$ values of -98 and -125 meV. These $J$ values significantly exceed the Landauer limit for minimum energy dissipation (18 meV), indicating their potential as promising candidates for spin logic operations at room temperature \cite{Landauer.1961,Lambson.2011}. Additionally, the magnetic coupling $J$ decreases as the spacer lengthens from direct linkage to -CCCC- and -Ph- spacers, arising from the reduction in the overlap between localized spin-orbitals (Fig. ~\ref{fig:fig4}e and S17-S26).

Besides Ovchinnikov’s rule, for a spin-polarized system, the Goodenough-Kanamori rule can be employed to predict and understand the magnetic interactions by analyzing the angle and type of adjacent spin-polarized orbitals. Generally, within the flat-bands, the limited dispersion ($t \approx 0$) will neutralize superexchange interaction ($-4t^2/U$), resulting in "orthogonal" extended wavefunctions \cite{Mori.2016}. Therefore, the potential exchange, favoring parallel spin alignments, prevails, leading to flat-band ferromagnetism. Despite substantial observations of flat bands in diamagnetic kagome 2D polymers, reports on metal-free ferromagnetism are still very rare \cite{Anindya.2022,ortiz2022theory}.

\begin{table*}
\caption{\label{tab:table1}Magnetic properties of the triangulene-based 2D polymers, including magnetic coupling ($J$), the magnetic moment per unit cell ($M$), electronic coupling ($|t|$), on-site Coulomb repulsion ($U$), $U/t$, and spin-polarization energy ($\Delta E_{\text{spin}}$).}
\begin{ruledtabular}
\begin{tabular}{cccccccc}
 &Systems& $J \text{ (meV)}$ & $M \ (\mu B)$ & $U \text{ (eV)}$ & $|t| \text{ (eV)}$ & $U/t$ & $\Delta E_{\text{spin}} \text{ (eV)}$ \\ \hline
&[TRI]&-31.8&0.00&2.66&0.16&16.6&-2.17 \\
&[TRI(N)]&59.2&2.00&0.35&0.01&35.00&-0.09 \\
&[TRI(B)]&55.9&2.00&0.38&0.01&38.00&-0.09 \\
&[TRI(B)-TRI(N)]& -124.8&0.00&0.27&0.12&2.25&-0.07 \\
&[PLY] & -97.6 & 0.00 & 2.61 & 0.11 & 23.7&-1.04 \\
\end{tabular}
\end{ruledtabular}
\end{table*}

Different from half-filled triangulene 2D polymers, such as [TRI] and [PLY], employing electron-sufficient building units as in [TRI(N)] will introduce additional p-electrons to the systems and elevate the Fermi level to the flat bands, while maintaining the stoichiometric and charge neutrality of the system \cite{Jing.2018}. The corresponding high density of states near the Fermi level stimulates spontaneous spin-polarization, as demonstrated by the large $U/t$ ratio of 35, according to the Stoner model \cite{Stoner.1938} (Fig. ~\ref{fig:fig3}d and ~\ref{fig:fig3}e and Table ~\ref{tab:table1}). Furthermore, the FM state is energetically more stable than the AFM state with a $J$ value of 59 meV, resulting in parallel spin alignment (Table ~\ref{tab:table1} and Fig. ~\ref{fig:fig3}f). This metal-free FM and corresponding high-spin ground state appear to be beyond the scope of Ovchinnikov’s rule. However, this behavior can be understood from the Goodenough-Kanamori rule. Since the kinetic exchange $-4t^2/U$, stemming from the virtual hopping of antiparallel spins, is suppressed by the limited band dispersion of the flat band ($t = 0.01$ eV), the direct exchange (i.e., potential exchange) between neighboring parallel spins governs the magnetic interaction, leading to the overall FM coupling (Table ~\ref{tab:table1}). Similarly, 2D polymers made of the electron-deficient TRI(B) monomer, i.e. [TRI(B)], shift the Fermi level down to the flat band as the p-electron is diminished compared to [TRI] (Fig. ~\ref{fig:fig3}g). Consequently, [TRI(B)] also emerges as a 2D FM polymer with a magnetic coupling of 56 meV, arising from the reduced electronic coupling of 0.01 eV and kinetic exchange of -1 meV (Table ~\ref{tab:table1}).

The Curie temperatures $T_{\text{c}}$ for [TRI(N)] and [TRI(B)] are predicted using Monte Carlo (MC) simulations, demonstrating spontaneous FM phase transitions occurring at about 260 K and 250 K, respectively. This indicates the possibility of forming long-range parallel spin ordering near room temperature (Fig. ~\ref{fig:fig4}b and ~\ref{fig:fig3}c). Remarkably, these $T_{\text{c}}$ surpass many inorganic 2D van der Waals FM magnets, such as $\text{Cr}_{2}\text{Ge}_{2}\text{Te}_{6} (J \approx 4\ \text{ meV}, T_{\text{c}} = 66 \text{ K}$) and $\text{CrI}_{3} (J \approx 4 \text{ meV}, T_{\text{c}} = 45\ \text{ K}$) \cite{Huang.2017,Huang.2018,Gong.2017}, which can be ascribed to the significantly stronger magnetic coupling in these 2D polymers with spin-polarized flat bands . We notice the long-standing controversy about the existence of magnetism in low-dimensional systems \cite{Jiang.2021b}, especially for the metal-free systems featuring small spin-orbit coupling and magnetic anisotropic energy, as Mermin-Wagner theorem prevents the long-range magnetic order at any finite temperature as consequences of thermal fluctuations \cite{Mermin.1966}. However, long-range spin order has been observed even in 1D systems, such as monatomic metal chains of cobalt and zigzag edges of narrow graphene nanoribbon \cite {Magda.2014,Gambardella.2002}, in contrast to the Ising model. Recent research shows that, at finite system size that is over several millimeters and exceeds the scale of typical nanotechnology devices, magnetic ordering can be stabilized by short-ranged interactions without any magnetic anisotropy, although for infinite-size system, it will be disturbed at non-zero temperature \cite{Jenkins.2022}. Our findings demonstrate that the spin-polarized flat bands at the Fermi level and metal-free FM can be achieved by manipulation of topology and chemical components in the non-Kekulé 2D polymers without the need for external doping or high magnetic fields.

In contrast to the AFM Mott insulators of carbon-centered triangulene 2D polymers and the semi-metal behavior of graphene, both [TRI(N)] and [TRI(B)] distinctly exhibit FM half-metallic properties, where one spin channel is (semi-)metallic and the other is single-band semiconductor (Fig. ~\ref{fig:fig3}e, ~\ref{fig:fig3}h, ~\ref{fig:fig4}a and ~\ref{fig:fig4}b). For [TRI(N)], the Fermi level intersects the Dirac point of the spin-down channel, indicating massless down-spin electrons, where the wavefunction is fully extended (Fig. ~\ref{fig:fig3}e, ~\ref{fig:fig4}a and S27). As the FM breaks the time-reversal symmetry and two spin-polarized bands have a linear crossing, these 2D polymers should host Weyl fermions in comparison to Dirac fermions that typically have four-component spinors and time-reversal symmetry \cite{Yan.2017}. Further validation of the emergence of Weyl fermions will be investigated in future work. Conversely, the spin-up channel shows unusual spin-polarized single-band semiconducting characteristics with a bandgap of 0.96 eV (Fig. ~\ref{fig:fig3}e and ~\ref{fig:fig4}a). This single-band feature provides an opportunity to disentangle the motions of charge carriers, owning to the mobile electrons at the parabolic conduction band maximum (CBM) and the heavy holes at the flat valence band. [TRI(B)] also demonstrates a semi-metallic feature for the spin-up channel, while the spin-down channel is semiconducting with a bandgap of 0.94 eV (Fig.~\ref{fig:fig3}h and ~\ref{fig:fig4}b). These characteristics imply that [TRI(N)] and [TRI(B)] can potentially serve as spin filters in organic spintronics for down-spin and up-spin electrons, respectively \cite{Ouardi.2013,Wang.2008}. Moreover, due to the linear dispersion at the Dirac point, spin-polarized currents can be transported with remarkably high mobility. Unlike inorganic systems containing heavy elements, organic materials composed of B, C, and N atoms generally possess smaller spin-orbit coupling, which impedes spin relaxation, thereby maintaining spin polarization over a larger diffusion length and longer relaxation time \cite{Dai.2018,Yu.2012,Alcon.2023}.

The binary 2D polymer, [TRI(B)-TRI(N)], composed of alternating nitrogen- and boron-centered triangulene monomers, is also half-filling similar to its isoelectronic counterparts [TRI] (Fig.~\ref{fig:fig3}j). However, distinct from [TRI], the different chemical potential in TRI(B) and TRI(N) induces a bandgap opening of 0.12 eV at the Dirac point for the diamagnetic state (Fig.~\ref{fig:fig3}j). This behavior parallels that observed in hexagonal boron nitride (hBN) relative to graphene and can be recapitulated by the TB model with an on-site energy difference of 0.1 eV (Fig.~\ref{fig:fig3}j). [TRI(B)-TRI(N)] also exhibits an AFM ground state as per Ovchinnikov’s rule analogous to [TRI] (Table ~\ref{tab:table1} and Fig.~\ref{fig:fig3}l). Interestingly, different from most AFM 2D polymers with degenerate spin-up and spin-down bands, for example, [TRI], [PLY], PTM-COF, TOT-COF, and TAM-COF \cite{Thomas.2019}, [TRI(B)-TRI(N)] is a spin-polarized half-semiconductor. In particular, [TRI(B)-TRI(N)] shows a narrow bandgap of 0.037 eV for spin-down electrons and a substantially larger one of 0.81 eV for spin-up electrons (Fig.~\ref{fig:fig3}k). This unprecedented half-semiconductor property in 2D polymers offers immense possibilities for controlling spin conductance through external fields. For instance, the application of laser light of a specific wavelength or an electric field could lead to the generation of a spin-polarized current via excitation or injection.

In summary, we proposed a spin-polarized flat band approach to realize metal-free ferromagnetism in 2D polymers. Using first-principles calculations, we examined 15 triangulene-based 2D polymers, where a distinct sandwiched electronic structure with a Dirac point and twin flat bands was unveiled. We demonstrated that half-filled systems, such as [TRI], [PLY], and [TRI(B)-TRI(N)], had AFM magnetic couplings of -32, -98 and -125 meV, respectively. Beyond half-filling, [TRI(N)] and [TRI(B)], showed metal-free FM with $J$ values of 59 and 56 meV. These FM couplings were due to the intersection of the Fermi level and the flat band, suppressing the electronic coupling and the AFM kinetic exchange, as per the Goodenough-Kanamori rule. MC simulations predicted Curie temperatures of 260 K and 250 K for these systems, implying the stability of long-range spin ordering at near-room temperatures. Compared to the AFM Mott insulators of [TRI] and [PLY], [TRI(N)] and [TRI(B)] were half-metals. In these 2D polymers, one spin channel was semi-metallic owing to the linear band crossing at the Fermi level, while the other spin channel was a single-band semiconductor. Additionally, we presented unprecedented AFM half-semiconductor properties in [TRI(B)-TRI(N)], where spin-polarized bands are energetically non-degenerate. These spin-dependent band gaps furnish considerable opportunities for the fine-tuning and control of spin-conducting behavior via external fields. Our findings not only enrich the fundamental understanding of the electronic and magnetic properties of 2D polymers, but also provide promising candidates for the realization of metal-free ferromagnetism and advancement of organic spintronics, particularly as spin filters or generators.

\begin{acknowledgments}
This project has been funded by the Alexander von Humboldt Foundation and Deutsche Forschungsgemeinschaft within CRC 1415 and RTG 2861. The authors would like to thank Dr. Thomas Brumme, Dr. Rico Friedrich, Dr. Ji Ma, Tsai-Jung Liu, Fupeng Wu and Xingyi Wang for helpful discussions. The authors gratefully acknowledge the computing time provided to them on the high-performance computers Noctua 2 at the NHR Center PC2. These are funded by the Federal Ministry of Education and Research and the state governments participating on the basis of the resolutions of the GWK for the national high-performance computing at universities.
\end{acknowledgments}

\nocite{*}

\end{document}